\documentclass[review]{elsarticle}

\usepackage{lineno,hyperref}
\usepackage{amsmath, multirow}

\journal{Nuclear Instruments and Methods in Physics Research Section A}

\begin{document}

\begin{frontmatter}

\title{Improvement in light collection of a photomultiplier tube using a wavelength-shifting plate}

\author[ucb]{Austin Mullen\corref{mycorrespondingauthor}}
\cortext[mycorrespondingauthor]{Corresponding author}
\ead{austin_mullen@berkeley.edu}
\author[LLNL]{Oluwatomi Akindele}
\author[LLNL]{Marc Bergevin}
\author[LLNL]{Adam Bernstein}
\author[LLNL]{Steven Dazeley}

\address[ucb]{University of California, Berkeley, Department of Nuclear Engineering, 4153 Etcheverry Hall, MC 1730, University of California, Berkeley, California 94720}
\address[LLNL]{Lawrence Livermore National Laboratory, 7000 East Avenue, Livermore, CA 94550}

\begin{abstract}
Large-volume water-Cherenkov neutrino detectors are a light-starved environment, as each interaction produces only $\sim 50-100$ photons per MeV. As such, maximizing the light collection efficiency of the detector is vital to performance. Since Cherenkov emission is heavily weighted towards the near UV, one method to maximize overall detector light collection without increasing the number of photomultiplier tubes is to couple each tube to a wavelength-shifting plastic plate, thus shifting photon wavelengths to a regime better suited to maximize photomultiplier efficiency and potentially detecting photons that miss the photocathode. To better understand the behavior of such plates, a scan of a rectangular wavelength-shifting plate was performed, and the results were used to calculate the overall percentage improvement in light collection that could be expected for individual PMTs in a large water-Cherenkov detector.  Measurements of a 15.1 in. by 11.5 in. wavelength-shifting plate using a 365 nm LED were found to increase overall light collection at the photomultiplier tube by $7.4\pm0.7\%$. A simulation tuned to reproduce these results was used to predict the behavior of a wavelength shifting plate exposed to Cherenkov spectrum light and found increases in light collection that were linear with edge length, assuming square geometries.  These results demonstrate the potential of wavelength-shifting plates to increase the overall light collection efficiency in a large detector.
\end{abstract}


\end{frontmatter}


\section{Introduction} 
The performance of a large-volume antineutrino detector, such as the AIT-NEO detector under development by the WATCHMAN Collaboration, depends in large part on its ability to collect the photons produced by antineutrino interactions at a few MeV within the detector volume \cite{bernstein-nonpro},\cite{bernstein2019ait}.  For this reason, maximizing the effective photocathode coverage of the photomultiplier tubes (PMTs) is desired.  One (expensive) method to do this is by increasing the total number of PMTs in the detector.  An additional way is to maximize the collection of near UV photons that miss the PMTs by using wavelength-shifting (WLS) plates \cite{johnston-WLSP}.

WLS plates are generally manufactured from a plastic material with a fluor added to absorb photons in the near UV, and re-emit them at a different, more favorable wavelength for detection \cite{eljen_site}.  The plates examined in this study, for example, were based on polyvinyl toluene (PVT) plastic, which absorb photons in the near UV range, with a maximum absorption at 355 nm, and re-emit them in the visible blue spectrum, with a maximum emission at 425 nm \cite{eljenspecs}.  WLS plates increase the light collection of a PMT in two ways.  First, the ability of a PMT to convert a photon into an electric signal depends heavily on the wavelength of the incoming photon, as determined by the quantum efficiency of the PMT \cite{h_bible}.  A WLS plate can shift photons from a wavelength where the PMT has a lower quantum efficiency to a wavelength of relatively higher sensitivity.  Second, photons absorbed in the plate are then re-emitted isotropically by the fluor.  Some of the light that would have originally missed the photocathhode of a PMT entirely can be re-emitted perpendicularly to its original direction of travel towards the photocathode.  Total internal reflection (TIR) serves to retain re-emitted photons in the plate, further enhancing this effect.

To better understand the behavior of WLS plates in a large-scale water-Cherenkov detector, it is important to measure and model the behavior of a single plate coupled to a PMT.  A light collection efficiency testing rig was developed at Lawrence Livermore National Laboratory for this purpose.  WLS plate performance was measured with this rig in air, and these measurements were used to build a Monte Carlo simulation.  The simulation was then used to predict the performance of larger, square plates exposed to Cherenkov spectrum light in both air and water, as plate behavior in water is expected to differ from that in air.  These differences are discussed later in this article, and are the study of a forthcoming empirical study of WLS plates in water.

\section{Experimental Design}
\begin{figure}
    \centering
    \includegraphics[width=0.8\linewidth]{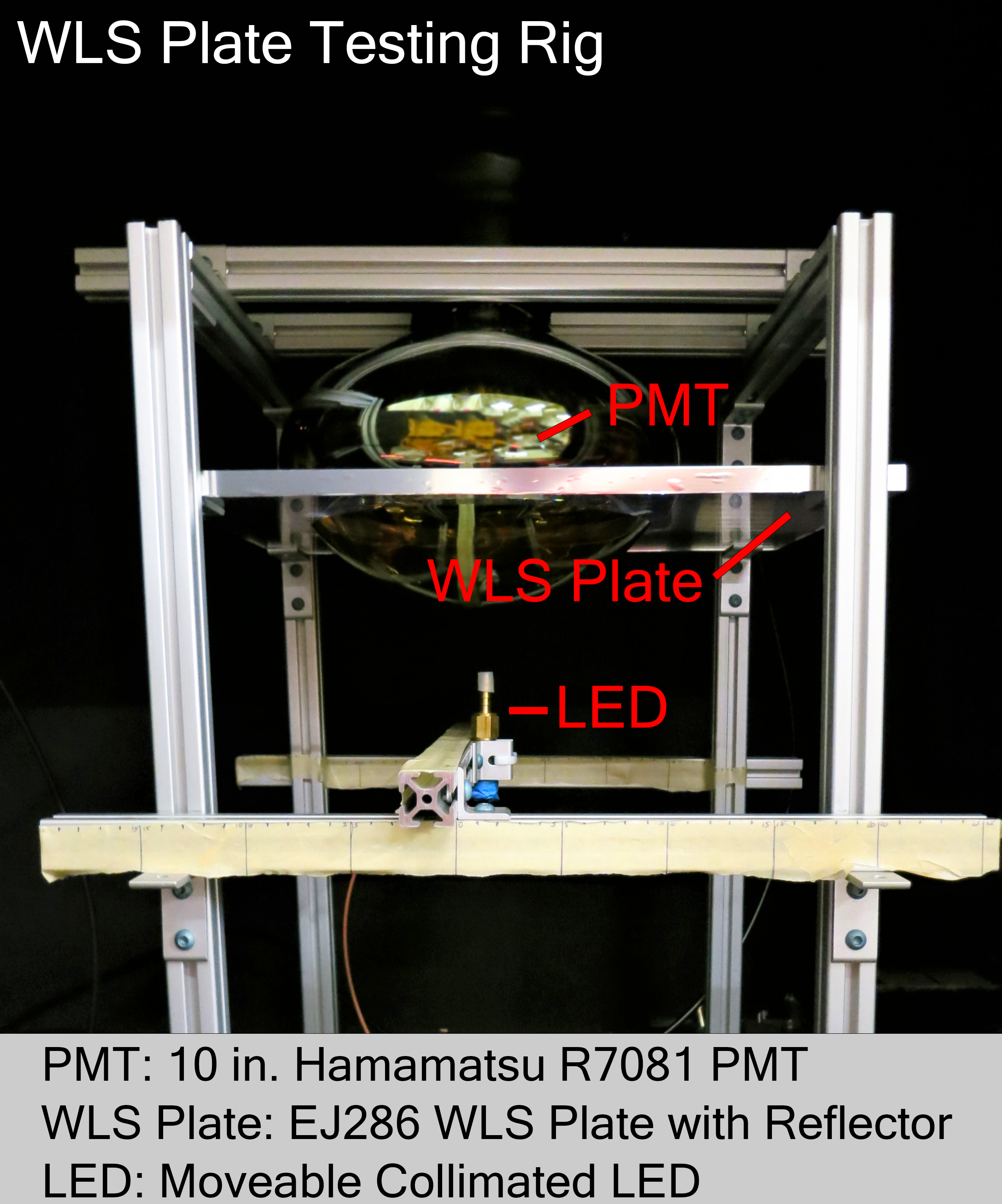}
    \caption{The light collection efficiency testing rig situated in a dark box, as seen from the side. The PMT sits face down on the hole in the WLS plate. The outer edge of the plate is wrapped in an aluminized mylar tape reflector so that light trapped within the plate is reflected back towards the PMT photocathode.  
    }
    \label{fig:dbfig}
\end{figure}

\begin{figure}
    \centering
    \includegraphics[width=\linewidth]{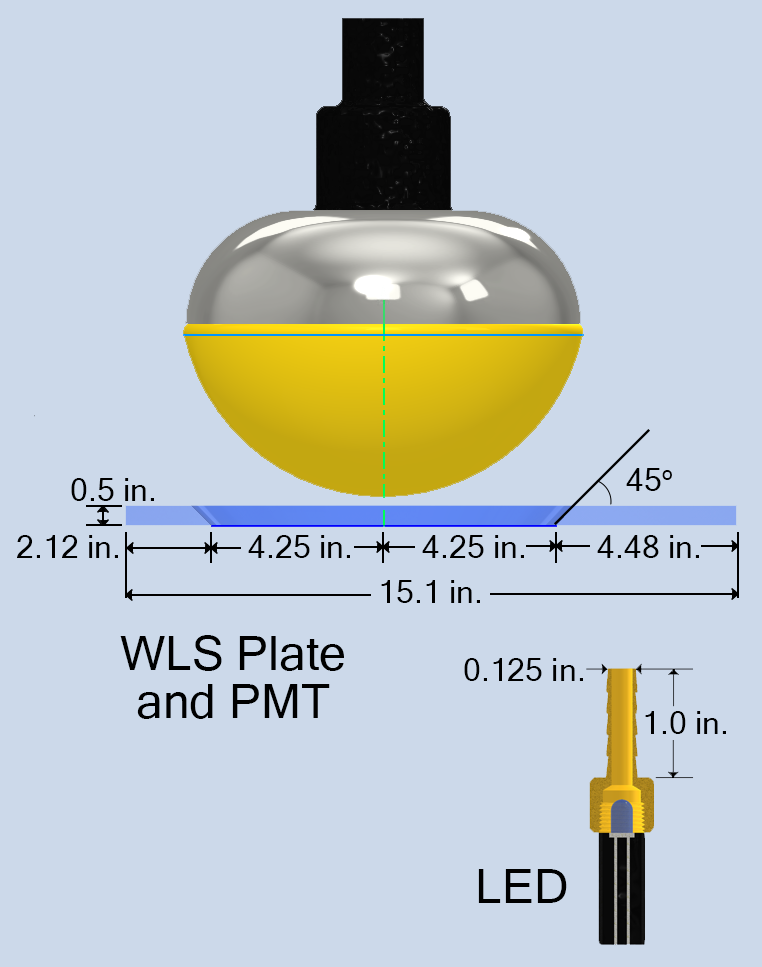}
    \caption{A computer aided drawing of the WLS plate, PMT, and LED collimator.  The edges of the hole in the WLS plate are beveled so that gravity will couple the PMT to the plate.  The collimator produces a spot with a diameter of 10 mm on the plate's surface.  }
    \label{fig:acfig}
\end{figure}

The testing rig developed for this experiment can be seen in Figure \ref{fig:dbfig}. A computer aided drawing of the rig, showing how the WLS plate and PMT are coupled, is given in Figure \ref{fig:acfig}.  A Hamamatsu R7081 253 mm diameter PMT is shown suspended in the air. It is coupled onto an Eljen Technology EJ-286 blue-emitting wavelength-shifting plastic plate \cite{hamamatsu-10in}.  This PMT was chosen as it is designed for use in large volume water-Cherenkov detectors, has a large surface area, and is currently the main candidate for use in the AIT-NEO detector.  The high voltage of the PMT was applied using a CAEN DT830X series desktop high voltage unit.  The output of the PMT was sent to a CAEN V1720 series digitizer to be converted directly into a digital pulse.  The coupling between the plate and PMT is provided by gravity with intimate points of contact along the interface.  The experiment was performed with and without optical grease (EJ-550) to improve the coupling between the plate and the PMT \cite{eljen_grease}.  While optical grease would not be able to be used in a real detector (as the PMT and plate will be submerged under water), experimental measurements with the optical grease were helpful to tune the behavior of the interface between the plate and the PMT in simulations.

The plate, as depicted in Figure \ref{fig:acfig}, is 15.1 in. (383.54 mm) in length and 11.52 in. (292.61 mm) in width, with a hole for the PMT offset from the center point of the plate by 1.05 in. (26.67 mm) in its longer dimension.  The wall of the hole is angled at 45 degrees, its smaller diameter being 8.5 in. (215.9 mm) and its larger diameter being 9.5 in. (241.3 mm).  The plate is 0.5 in. (12.7 mm) thick.  The rectangular, asymmetric shape of the WLS plate was a design choice to allow for its inclusion in a future testing campaign in a one-ton water-Cherenkov detector, which will be briefly discussed later in this article.

A collimated LED, also depicted in Figure \ref{fig:acfig} was placed on a movable rail system 2.5 in. (63.5 mm) beneath the surface of the WLS plate.  It was placed at this distance to ensure that the collimator did not collide with the bulb of the PMT during testing.  The LED produced a spot with a diameter of 10 mm at the plate's surface, and measurements were taken at 40 mm intervals along the short side of the plate and 10 mm intervals along the long side of the plate.  The LED can be easily removed from its housing and replaced to allow for tests with different LEDs.  Two LEDs were used - a 365 nm UV LED and a 569 nm green LED. The green LED was chosen as a control, since green light is not absorbed in the WLS plate.  The entire rig was constructed in a dark box painted with non-reflective black paint to minimize the amount of stray reflected light from the box's surfaces.
Reflective aluminized milar tape was placed around the outer edge of the WLS plate to ensure that the light captured by the plate would be reflected back towards the PMT. The tape is highly reflective (with a reflectivity of around 95\%) and specular.

\section{Results}
\begin{figure}
    \centering
    \includegraphics[width=\linewidth]{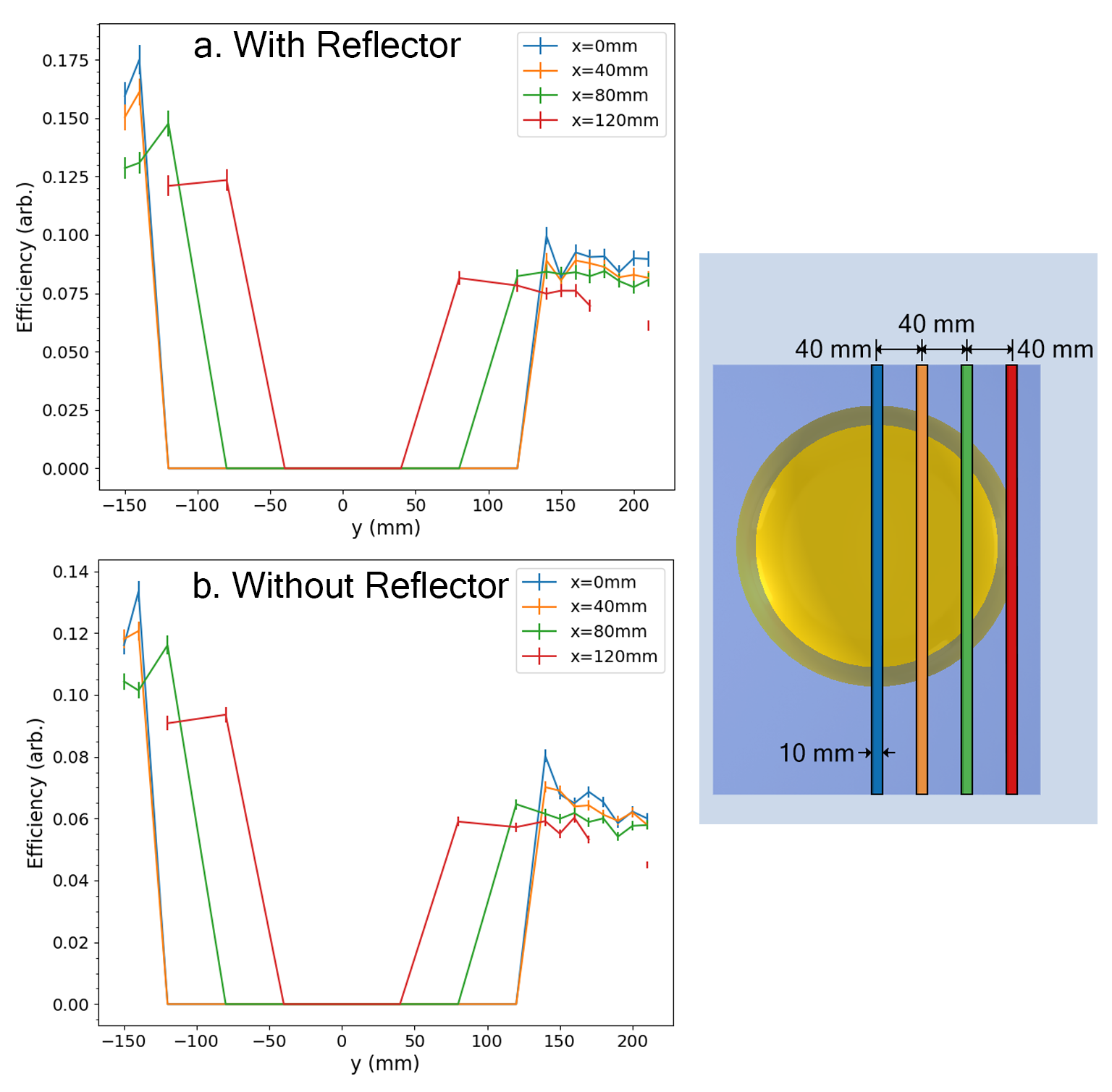}
    \caption{Various cuts lengthwise across the plate which show the variation in the light collection efficiency over the surface of a WLS plate with (a.) and without (b.) an aluminized tape reflector.  Measurements on the bulb of the PMT itself are excluded, along with measurements that were blocked by the structural brackets.  The colors of each data set can be seen projected onto a bottom view of the PMT plate system in the image inset to the right.  The blue data is for the center line of the WLS plate, with the other data collected at 40 mm intervals outwards from that line.  In both cases, there is a marked assymetry between the positive and the negative points due to the asymmetric shape of the WLS plate used.  Additionally, both cases demonstrate that light collection is largely constant across a majority of the plate's surface, from approximately 150 to 220 mm.}
    \label{fig:1d}
\end{figure}

Data was collected in such a way as to measure the change in light collection efficiency across different points on the plate using the UV LED.  Figure \ref{fig:1d} shows a comparison of light collection efficiency over the surface of the plate, both with and without a reflector. The efficiencies presented were measured relative to  that at the center of the PMT bulb. The efficiency was found to be roughly constant over the plate's surface, except very near to the PMT-plate boundary, where the LED spot size was found to have a significant effect.

An asymmetry in the light collection efficiency between the two sides of the WLS plate was observed; in Figure \ref{fig:1d}, the negative y direction produced a higher collection efficiency than those in the positive y direction.  This asymmetry was likely driven by the shorter distance between the outer edge of the plate and the PMT in the negative y direction. This means that photons emitted on the narrower side of the plate have shorter average path lengths to reach the PMT than those emitted in the longer side of the plate.  This, in turn, means that fewer photons are lost to re-absorption in the plastic.  Once reabsorbed, photons can be either lost to non-radiative processes or re-emitted at a direction not amenable to capture in the plate by TIR. A similar, but smaller, asymmetry was observed in the simulation results. The plate was also simulated using very long photon attenuation lengths, which caused the asymmetric response to disappear.  This indicates that re-absorption due to longer photon path lengths in the plastic is likely the driving cause of the asymmetric response. A symmetric plate was also simulated,
and the asymmetry again disappeared.

A comparison of light collected with and without the WLS plate is presented in Table \ref{tab:sim-comp-eff}, along with the results from the Monte Carlo simulations that will be discussed in the next section.  The increase in light collection is defined here as the fractional increase (in percent) in the number of photoelectrons generated in an event between a PMT with a wavelength shifting plate and a PMT without.  This is given in Equation 1, where $C$ is the light collection increase and $L$ is the number of photoelectrons generated by the PMT.  To calculate the light collection of the PMT without the WLS plate, a green LED was used, as the plate is mostly transparent to it.  This allows the green LED measurements to serve as a proxy for a measurement taken without the WLS plate.  The normalization to the response at the center of the PMT bulb corrects for the PMT's different quantum efficiencies at the two wavelengths.

\begin{equation}
    C = \frac{L_{WLS}-L_{Bare}}{L_{Bare}}
\end{equation}

It is notable in Table \ref{tab:sim-comp-eff} that the WLS plate without a reflector provides no statistically significant benefit over a bare PMT in this experimental setup.  This is due to the lower overall light collection efficiency of a plate without a reflector, combined with a shadowing effect where light that would otherwise have struck the PMT surface is instead absorbed by the plate and directed away from the PMT.  While this shadowing effect is present with a reflector as well, the overall light collection efficiency is much higher, reducing its impact.  It is also notable that coupling the PMT to the WLS plate with optical grease roughly doubles the light collection increase compared to air.  This is consistent with simple optics calculations that can be performed to show that, in air, roughly half of the collected light will be rejected from the plate-PMT interface due to TIR.  

\begin{table}
    \centering
    \begin{tabular}{c||c|c}
         \multirow{2}{*}{Reflector}& \multicolumn{2}{c}{Light Increase} \\\cline{2-3}
        & Experimental             & Simulated\\
        \hline
        Yes & 7.4 $\pm$ 0.7\% & 7.1 $\pm$ 0.2\%\\
        No & -0.08 $\pm$ 0.7\% & -1.2 $\pm$ 0.2\%\\
        Yes \& using Optical Grease & 16.2 $\pm$ 0.7\% & 16.2 $\pm$ 0.2\%\\
    \end{tabular}
    \caption{The relative increase in light detected from a WLS plate coupled with a PMT. The simulation replicates the observed light increase.  The calculated increases assume conditions of uniform irradiation of the plate and PMT.}
    \label{tab:sim-comp-eff}
\end{table}

\section{Comparison to Simulations}
\begin{figure}
    \centering
    \includegraphics[width=\linewidth]{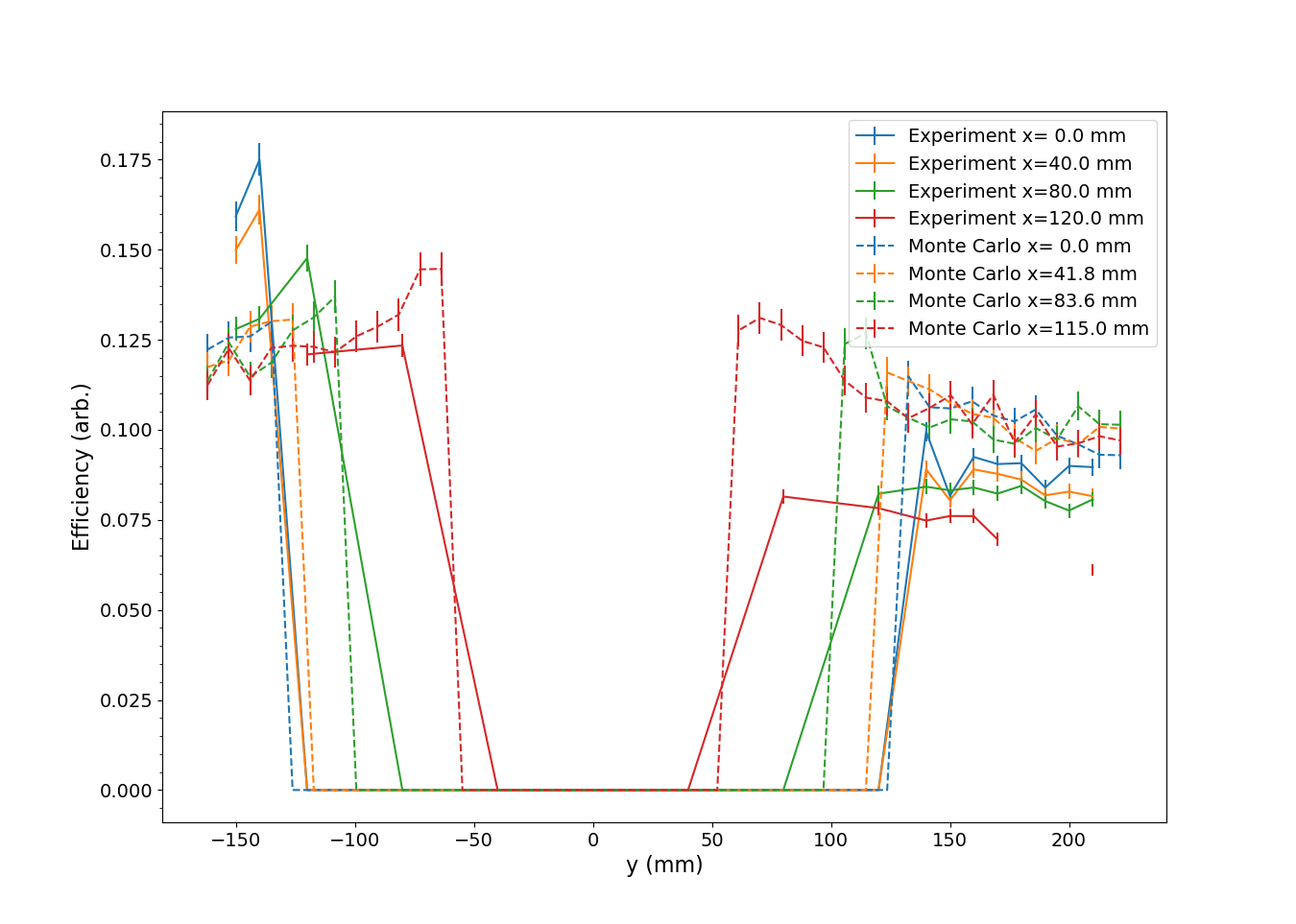}
    \caption{The light collection efficiency from both the experimental and simulation results across the plate's surface.  Both the experimental and simulated results are largely consistent, though the simulations do not recreate the asymmetry to the extent seen in the experimental results.}
    \label{fig:sim-comp}
\end{figure}

In order to better generalize these results to plate sizes and geometries not tested here, or conditions beyond those possible in the dark box, a simulation was created in the Monte Carlo framework RAT-PAC (Reactor Analysis Tool, Plus Additional Codes), which is a wrapper for Geant4 \cite{rat-pac}.  The plate and PMT used were recreated in the simulation according to manufacturer specifications. The properties of the plate, such as the roughness of the plate's surfaces, were then tuned to bring its behavior in the simulations into better agreement with its behavior in the experimental setup.  A comparison between the experimentally measured light collection efficiency and a simulated efficiency across a plate's surface is given in Figure \ref{fig:sim-comp}.  It can be seen that the general behavior of the light across the wavelength shifting plate is represented in the simulation, and some amount of asymmetry is also observed, though not to the same extent seen in the experimental results.  The overall increases in light collection can be calculated from simulation results, as shown in Table \ref{tab:sim-comp-eff}.  The given uncertainties are statistical from the Monte Carlo simulation.  The results are largely consistent between the experiment and the simulation for all three experiments.

In a real detector the plates and PMTs will be submerged under water rather than suspended in air.  The two media will behave differently due to their distinct indices of refraction leading to different reflection and refraction properties at the plate surfaces.  Simulations of a 15.1 in. by 11.52 in. plate identical to the one tested experimentally here were conducted with the surrounding air changed to water.  The simulations resulted in a light collection increase in water of $6.7\pm0.2\%$, a small reduction relative to the same plate in air.

\begin{figure}
    \centering
    \includegraphics[width=\linewidth]{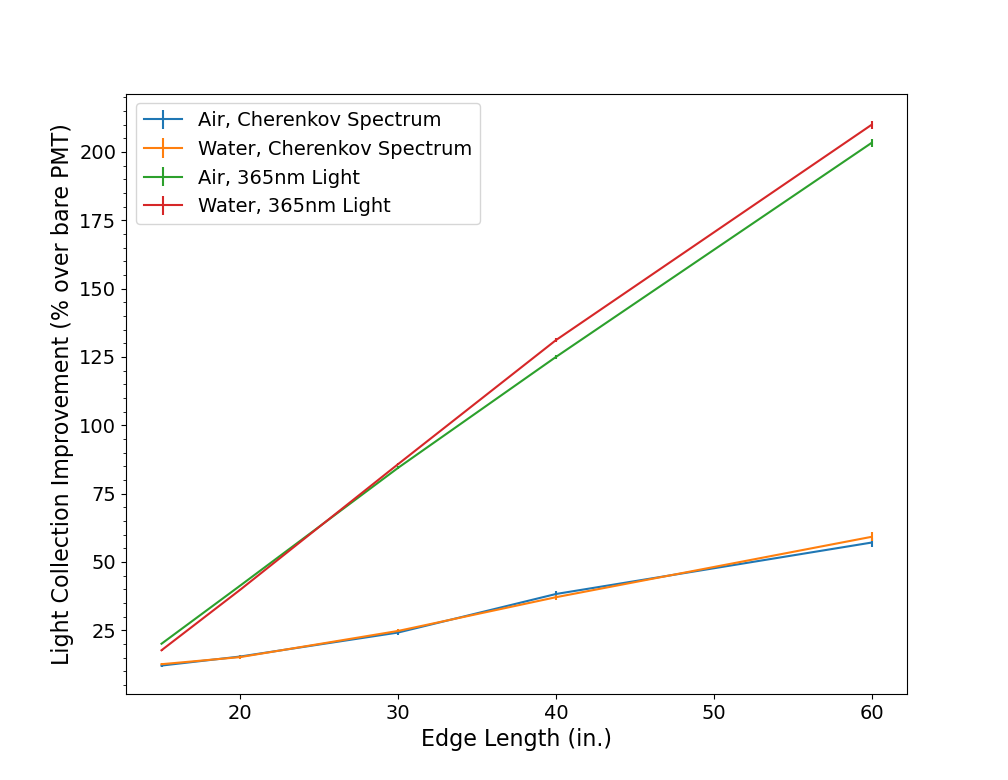}
    \caption{Simulation results for the light collection increases afforded by various square plate sizes in air and water.  The relationship between edge size and light collection improvement is seen to be linear.  The ability of the WLS plate to collect Cherenkov spectrum light is seen to be reduced compared to its ability to collect monochromatic 365 nm light.}
    \label{fig:large_series}
\end{figure}

Finally, it is of interest to test, via simulations, the increase in total light collection of a larger plate with a more regular, square geometry, such as one that may be included in a large-volume water-Cherenkov antineutrino detector. The simulations were run using both uniform 365 nm light, such as that produced by the LED in the experimental setup, and the diffuse Cherenkov spectrum produced in a real detector.  The results, for both plates in air and in water, are given in Figure \ref{fig:large_series}.  It can be seen that over a wide range of WLS plate sizes, the light collection increase scales approximately linearly with plate edge length, and can be described empirically by Equations 2-5, where $x$ is the side length of a square plate and $C$ is the improvement in light collection afforded by the WLS plate.  The equations should only be taken to be valid under the limited range of plate sizes tested here and not extrapolated to radically different plate sizes, as they fail to accurately represent the behavior of very small plates where x approaches zero, or very large plates, which begin to behave non-linearly.  The linear relationship observed can be interpreted in the following way -  while the total light incident on the plate's surface increases as roughly $x^2$, the absolute collection efficiency of the plate falls off as $1/x$.

\begin{align}
    \textrm{For Air, Cherenkov Spectrum: } C &= (1.00 \pm 0.03)x - (3.07 \pm 0.75)\\
    \textrm{For Water, Cherenkov Spectrum: } C &= (1.00 \pm 0.03)x - (3.56 \pm 0.73)\\
    \textrm{For Air, 365 nm Light: } C &= (4.18 \pm 0.02)x - (42.44 \pm 0.39)\\
    \textrm{For Water, 365 nm Light: } C &= (4.45 \pm 0.03)x - (48.91 \pm 0.73)
\end{align}

It should be noted that the overall improvement in light collection is much poorer when considering Cherenkov spectrum light compared to monochromatic 365 nm light, with Cherenkov light resulting in a smaller improvement relative to the monochromatic source (16\% versus 40\% for a 20 inch plate in water, for example).  This is because the WLS plate is transparent to many of the longer wavelengths in the Cherenkov spectrum, rendering the plate ineffective at collecting this light.  Additionally, the WLS plate is most effective at collecting UV light with very short wavelengths.  However, this light (particularly light with a wavelength of less than 300 nm) is unlikely to reach the plate at all, as it has a very short attenuation length in water.  This effect biases the spectrum to which the WLS plate is exposed towards longer-wavelength light, which the plate is less effective at collecting.  This further drives the relative reduction of efficiency of a plate exposed to Cherenkov spectrum light compared to one exposed to monocromatic light.

One can see broad consistency between the results for both the simulations performed in air and in water.  It is possible that this consistency is due to two competing factors that change with the medium.  First, more light will escape from the plate's surfaces because the indices of refraction between the plate and water are more similar than those of the plate and air, meaning that TIR is less effective at retaining light in the plate.  Second, once light is captured by the plate, a greater fraction may be able to cross from the plate into the PMT itself, as, for the same reason, less light will be rejected from the boundary by TIR. Our results indicate that these two countervailing effects roughly cancel each other.  More study is still needed, however, to experimentally test the behavior of a WLS plate in water and confirm these effects.  One such study will be carried out by testing the light collection of four WLS plates submerged underwater in a one-ton water-Cherenkov detector. Results from this experiment will be presented separately.

The results presented in this work are partially consistent with previous results in the literature.  In a similar test of a single plate and PMT system, Johnston calculates that a 20 inch circular WLS plate of equivalent plastic exposed to a Cherenkov spectrum will produce a 13.1\% increase in light collection compared to a bare PMT \cite{johnston-WLSP}. This is consistent with the expected light improvement of a square plate of equivalent surface area calculated here, which has an edge length of 17.8 inches and an expected light collection improvement of 14\%.  However, when simulating 27 inch square plates in a large water-Cherenkov detector, Johnston finds a higher light collection improvement of 31\% for PMTs with WLS plates compared to those without.  This is less consistent with the findings here, which expects an improvement of only 23\% for a 27 inch square plate exposed to a Cherenkov spectrum in water.  Factors such as the degree of optical coupling between the plate and the PMT and the reflectivity of the reflector employed on the plate may drive this difference.  The PMTs in Johnston's simulations were also assumed to have a uniform collection efficiency across the bulb, while the efficiency of the PMTs simulated in this work was reduced on the outer edges of the bulb to better match real PMT performance.  This difference in simulated PMT behavior may also contribute to the observed difference in light collection.

\section{Conclusions}
In this experiment we have demonstrated that a small, rectangular wavelength-shifting plastic plate, such as the one measured here, can increase the overall light collection of a PMT by $7.4\pm0.7$\% based on measurements performed with a 365 nm UV LED.  These measurements were then replicated in a Geant4 simulation of the plate.  The simulations suggest that WLS plates are scalable, as larger simulated plate geometries demonstrated an approximately linear relationship between increased plate edge size and improvements in light collection when exposed to Cherenkov spectrum light.  While this work has investigated the effects of wavelength-shifting plates on light collection, other possible effects, such as those on the timing profile of the light arriving at the PMT and possible cross-talk caused by light scattering between different PMTs, will also be important when considering the overall effects that the inclusion of WLS plates may have on a large scale water-Cherenkov antineutrino detectors.  These effects should be investigated in future studies.  Additionally, the light collection efficiency of the plates in water instead of air should also be studied experimentally to better understand actual plate behavior in a water-filled detector.

\section*{Acknowledgements}
This work was performed under the auspices of the U.S. Department of Energy by Lawrence Livermore National Laboratory (LLNL) under Contract DE-AC52-07NA27344, release number LLNL-JRNL-831975. This material is based upon work supported by the Department of Energy National Nuclear
Security Administration through the Nuclear Science and Security Consortium under Award Number(s) DE-NA0003180 and/or DE-NA0000979.
This report was prepared as an account of work sponsored by an agency of the United States
Government. Neither the United States Government nor any agency thereof, nor any of their
employees, makes any warranty, express or limited, or assumes any legal liability or
responsibility for the accuracy, completeness, or usefulness of any information, apparatus,
product, or process disclosed, or represents that its use would not infringe privately owned
rights. Reference herein to any specific commercial product, process, or service by trade name,
trademark, manufacturer, or otherwise does not necessarily constitute or imply its
endorsement, recommendation, or favoring by the United States Government or any agency
thereof. The views and opinions of authors expressed herein do not necessarily state or reflect
those of the United States Government or any agency thereof.

\bibliography{mybibfile}

\end{document}